\documentclass[aps,prd,twocolumn,nofootinbib]{revtex4-2}

\usepackage{booktabs,makecell,multirow}

\usepackage{graphicx}
\usepackage{slashed}
\usepackage{amsfonts}
\usepackage{mathrsfs}
\usepackage{amsmath}
\usepackage{amssymb}
\usepackage{color}

\usepackage[colorlinks=true, linkcolor=blue, citecolor=blue, urlcolor=blue]{hyperref}

\usepackage[normalem]{ulem}

\usepackage{array}
\begin{document}

\title{Scheme-independent determination of the QCD running coupling at all scales\\ from jet observables using the principle of maximum conformality and infinite-order scale setting}

\author{Leonardo Di Giustino$^{1,2}$}
  \email[email:]{leonardo.digiustino@uninsubria.it}
\author{Stanley J. Brodsky$^{3}$ }
  \email[email:]{sjbth@slac.stanford.edu}
\author{Philip G. Ratcliffe$^{1,2}$ }
  \email[email:]{philip.ratcliffe@uninsubria.it}
\author{Sheng-Quan Wang$^{4}$}
  \email[email:]{sqwang@alu.cqu.edu.cn}
\author{Xing-Gang Wu$^5$}
  \email[email:]{wuxg@cqu.edu.cn}

\address{$^1$Department of Science and High Technology, University of Insubria, via Valleggio 11, I-22100, Como, Italy}
\address{$^2$INFN, Sezione di Milano--Bicocca, 20126 Milano, Italy}
\address{$^3$SLAC National Accelerator Laboratory, Stanford University, Stanford, California 94039, USA}
\address{$^4$Department of Physics, Guizhou Minzu University, Guiyang 550025, P.R. China}
\address{$^5$Department of Physics, Chongqing University, Chongqing 401331, P.R. China}

\date{\today}

\begin{abstract}
  We present a new approach to determining the strong coupling $\alpha_s(Q)$, over the entire range of validity of perturbative QCD, for scales above  $\Lambda_{\mathrm{QCD}}$ and up to the Planck scale $\sim1.22\cdot10^{19}$\,GeV, with the highest precision and using the data of a single experiment. In particular, we use the results obtained for the thrust ($T$) and $C$-parameter ($C$) distributions in $e^+e^-$ annihilation at a single annihilation energy $\sqrt{s}=M_Z$ (i.e.\ at the $Z^0$ peak). This new method is based on the \emph{intrinsic conformality} (iCF) and on the Infinite-Order Scale Setting, using the Principle of Maximum Conformality (i.e.\ the PMC$_\infty$), which allows a rigorous determination of the renormalization scales for the event-shape variable distributions satisfying all of the requirements of Renormalization Group Invariance, including renormalization-scheme independence and consistency with Abelian theory in the $N_C \to 0$ limit. This new method is based on the scale-invariance of the iCF, which allows determination of $\alpha_s(\mu_0)$ at any scale $\mu_0$, and on the Maximum Likelihood statistical approach.   
  
 We propose a novel approach to determining the best-fitting range by considering all possible intervals over the entire range of bins available in the perturbative region and selecting that which returns the most-likely-lowest $\chi^2_{\rm min}$.
  This new method is designed to eliminate the errors that arise due to selection of the bin-interval and that have been neglected in previous analyses. 
 In particular, using data for thrust and $C$-parameter at the $Z^0$ peak from ALEPH, OPAL, DELPHI and L3 experiments, we obtain the average value: $\alpha_s(M_Z)= 0.1182^{+0.0007}_{-0.0007}$, for the strong coupling.  This determination of $\alpha_s(M_Z)$ is consistent with the world average and has an improved precision with respect to the values obtained from the analysis of event shape observables currently used in the world average.

\end{abstract}

\maketitle

The QCD coupling $\alpha_s(Q)$ is a fundamental standard-model (SM) parameter; its determination to the highest precision is crucial for testing theory and revealing possible new-physics (NP) signals, which may be obscured by QCD dynamics. The conventional practice of taking the renormalization scale as that typical of the process ($\mu_r=Q$) and estimating the theoretical uncertainties via variation by factors of two ($Q/2\leq\mu_r\leq2 Q$) is afflicted with renormalization scale and scheme ambiguities. Such ambiguities originate from the truncation of the perturbation series and the incomplete cancelation of the scale dependence under the renormalization group equations (RGE) due to the missing higher-order coefficients in the pQCD expansion. While we do maintain that perturbative calculations to all orders are fundamental for theoretical predictions, they are not essential and may not truly resolve the scale and scheme ambiguities, given the nature of QCD and the asymptotic nature of the perturbation series. The inconsistencies of the conventional scale setting practice were demonstrated in our recent review Ref.~\cite{DiGiustino:2023jiq}.

In this article we shall demonstrate that a recently proposed method~\cite{DiGiustino:2020fbk} for solving the renormalization scale and scheme ambiguities, namely \emph{infinite-order scale setting using the principle of maximum conformality} (PMC$_\infty$), based on intrinsic conformality (iCF) and by using the PMC, leads to an improved extraction of $\alpha_s(Q)$ at the $Z^0$ peak from the thrust and $C$-parameter distributions in the annihilation process $e^+e^-\to3$\,jets. Moreover, our new results agree with the most recent world average given by the Particle Data Group~\cite{Navas:2024a1}, improving the precision and significantly reducing the theoretical errors related to the renormalization scale. 

In addition, we propose a new method to eliminate errors related to the choice of the bin-interval in the perturbative region, which is consistent with the Maximum-Likelihood approach and the $\chi^2$-minimization procedure used for the fit of the strong coupling.

The current world average in the $\overline{\mathrm{MS}}$ scheme, exploiting lattice QCD, is $\alpha_s(M_Z)=0.1180\pm0.0009$ and without lattice QCD, $\alpha_s(M_Z)= 0.1175\pm0.0010$~\cite{Navas:2024a1}. Other recent extractions of $\alpha_s(Q)$ from the same process using the conventional scale setting have obtained values either significantly discrepant with respect to the world average or affected by large theoretical uncertainties.

For example, Ref.~\cite{Dissertori:2009ik}, using NNLO+NLL predictions, gives $\alpha_s(M_Z)=0.1224\pm 0.0039$, with a perturbative uncertainty of $0.0035$; Refs.~\cite{Abbate:2010xh} and \cite{Hoang:2015hka} present extractions of $\alpha_s$ by matching resummation calculations up to N$^3$LL accuracy: $\alpha_s(M_Z)=0.1135\pm0.0011$ from thrust and $\alpha_s(M_Z)=0.1123\pm0.0015$ from the $C$-parameter respectively; further corrections, such as nonperturbative hadronization effects, are  introduced in Ref.~\cite{Dasgupta:2003iq}, although, as pointed out in \cite{Tanabashi:2018oca}, the theoretical uncertainties in extracting $\alpha_s$ using Monte Carlo generators to simulate hadronization effects are not well understood.

We shall demonstrate here that the PMC$_\infty$ based on the iCF allows determination of $\alpha_s(Q)$ at any scale, even up to very high values, where the theory might be expected to breakdown, whereas conventional scale setting allows only one value of $\alpha_s$ at the kinematic scale $\sqrt{s}$ of the process to be extracted (with errors given by the uncertainty in choosing the renormalization scale).

The conventional scale setting is not only affected by renormalization scale and scheme ambiguities, but it also violates renormalization group invariance (RGI) and leads to results that depend on arbitrary conventions (such as the renormalization scheme) and are affected by large theoretical errors. Moreover, estimation of the unknown higher-order terms by simply varying the renormalization scale within a range of two, is arbitrary and affected by factorially growing (``renormalon'') terms, $n!\,\beta^n_0\,\alpha^{n+1}_s$, occurring in perturbation theory and related to the $\beta$ terms. Consistency of the conventional scale setting with the Gell-Mann--Low procedure~\cite{GellMann:1954fq} for QED is also lacking. In fact, as was proposed in Ref.~\cite{Brodsky:1997jk}, the pQCD predictions should match the Abelian theory analytically in the $N_C\to0$ limit.

The principle of maximum conformality (PMC) \cite{Brodsky:2011ta, Brodsky:2012rj, Brodsky:2011ig, Mojaza:2012mf, Brodsky:2013vpa, Yan:2022foz} provides a systematic way to eliminate renormalization scheme and scale ambiguities. The PMC scales are set by reabsorbing the $\beta$ terms occurring in the perturbative calculation via the renormalization group equation (RGE). These terms reveal the subprocess quark and gluon QCD dynamics, which also governs the strength of the running coupling. Since PMC predictions are independent of the choice of renormalization scheme, PMC scale setting satisfies RGI~\cite{Brodsky:2012ms,Wu:2014iba,Wu:2019mky}. And, since the $\beta$ terms are reabsorbed into the PMC scales, the pQCD series also becomes renormalon free. The PMC method extends Brodsky--Lepage--Mackenzie (BLM) scale setting~\cite{Brodsky:1982gc} to all orders, all processes and all gauge-theories.

The PMC$_\infty$ was introduced in Ref. \cite{DiGiustino:2020fbk} and it is based on the \emph{Principle of Maximum Conformality} and it preserves the intrinsic conformality -(iCF).
The iCF is a Renormalization Group invariant parameterization of the perturbative observable characterized by conformal-scales and conformal-coefficients, that leads to a scale invariant quantity at all orders. In particular, in Ref. \cite{DiGiustino:2020fbk}, it was suggested this property to be well preserved in the event shape variables thrust and C-parameter at NNLO. By introducing the PMC strategy in this framework , i.e. reabsorbing the $\beta_0$-term, we obtain the conformal PMC$_\infty$-scales and the conformal PMC$_\infty$-coefficients. Once these coefficients are determined, they are extended to all orders following the strong coupling RG equation. Thus the PMC$_\infty$ is an \emph{infinite-order scale-setting} and preserves the features of the other PMC approaches \cite{Wu:2014iba}.

A precise determination of $\alpha_s(M_Z)$ can be obtained from event-shape observables in electron--positron annihilation~\cite{Kluth:2006bw}, in particular, to 3\,jets~\cite{Dissertori:2009ik} (since the LO process is already proportional to $\alpha_s(Q)$) by comparing the theoretical predictions with the experimental data at the $Z^0$ peak from the ALEPH, DELPHI, OPAL, L3 and SLD experiments~\cite{ALEPH:2003obs, DELPHI:2003yqh, OPAL:2004wof, L3:2004cdh, SLD:1994idb}.

Although extensive studies of these observables have been made over the last few decades, including higher-order corrections from next-to-leading order (NLO) calculations \cite{Ellis:1980wv, Kunszt:1980vt, Vermaseren:1980qz, Fabricius:1981sx, Giele:1991vf, Catani:1996jh} to the next-to-next-to-leading order (NNLO) \cite{Gehrmann-DeRidder:2014hxk, Gehrmann-DeRidder:2007nzq, GehrmannDeRidder:2007hr, Weinzierl:2008iv, Weinzierl:2009ms} and including resummation of the large logarithms \cite{Abbate:2010xh, Banfi:2014sua}, the theoretical predictions are still affected by significant uncertainties related to the large renormalization-scale ambiguities and for the case of 3-jet event-shape distributions, extracted values of $\alpha_s$ deviate from the world average~\cite{Navas:2024a1}.

The remaining parts of the paper are origanised as follows: in Sec. \ref{sec:alphas}, we introduce the strong coupling and the scale dependence; in Sec. \ref{sec:eventshapes}, we recall the event shape variables and the PMC$_\infty$ results; in Sec. \ref{sec:method}, we discuss the fitting method and the new approach for determining the best-fitting range; in Sec. \ref{sec:results}, we show and discuss the results; in Sec. \ref{sec:conclusions}, conclusions.

\section{The strong-coupling evolution}
\label{sec:alphas}
The evolution of the strong coupling $\alpha_s(\mu)$ is governed by the $\beta$-function, which is defined as:
\begin{align} 
  \beta(\alpha_s)&\equiv \frac{1}{4
  \pi}\frac{d\alpha_s(Q)}{d \log Q^2} = -\left(\frac{\alpha_{s}}{4
  \pi}\right)^{2} \sum_{n=0} \left(\frac{\alpha_{s}}{4
  \pi}\right)^{n} \beta_{n}. \label{betaf}
\end{align}
with the $\beta$-coefficients, $\beta_n$, $n=0,...,4$,  from one- to five-loop accuracy respectively, calculated in a given scheme. Results for these coefficients can be found in Refs.~\cite{Gross:1973id,Politzer:1973fx,Caswell:1974gg, Jones:1974mm,Egorian:1978zx, Larin:1993tp,vanRitbergen:1997va,Baikov:2016tgj}. For a review of the strong coupling and of all the $\beta$ coefficients, see e.g.~\cite{Deur:2016tte,Blumlein:2023aso}. In particular, we use the $\overline{\mathrm{MS}}$ scheme throughout this paper. For the running coupling $\alpha_s(Q)$, we use the solutions at 5-loop accuracy given by both the RunDec program \cite{Chetyrkin:2000yt} and by the perturbative solution given by
\begin{widetext}
\begin{align}{\label{5L-as}}
   a_{\mu} & =  a_{\mu_0}+\beta_0 \ln
  \left(\frac{\mu_0^2}{\mu^2}\right) a_{\mu_0}^2+\left[\beta_0^2
  \ln^2 \left(\frac{\mu_0^2}{\mu^2}\right)+\beta_1
  \ln\left(\frac{\mu_0^2}{\mu^2}\right) \right] a_{\mu_0}^3
\nonumber \\
  & \qquad +\left[\beta_0^3 \ln^3
  \left(\frac{\mu_0^2}{\mu^2}\right)+\tfrac{5}{2} \beta_0 \beta_1
  \ln^2\left(\frac{\mu_0^2}{\mu^2}\right)+\beta_2 \ln
  \left(\frac{\mu_0^2}{\mu^2}\right)\right] a_{\mu_0}^4
\nonumber \\
  & \qquad +\left[\beta_0^4 \ln^4\left(\frac{\mu_0^2}{\mu^2}\right)
  +\tfrac{13}{3} \beta_0^2 \beta_1
  \ln^3\left(\frac{\mu_0^2}{\mu^2}\right)+\tfrac{3}{2} \beta_1^2
  \ln^2\left(\frac{\mu_0^2}{\mu^2}\right)+3 \beta_2 \beta_0
  \ln^2\left(\frac{\mu_0^2}{\mu^2}\right)+\beta_3 \ln
  \left(\frac{\mu_0^2}{\mu^2}\right)\right] a_{\mu_0}^5
\nonumber \\
  & \qquad +\left[  \beta_0^5 \ln^5 \left(\frac{\mu_0^2}{\mu^2}\right)
  +\tfrac{77}{12}\beta_1\beta_0^3
  \ln^4\left(\frac{\mu_0^2}{\mu^2}\right)+ \left(6 \beta_2\beta_0^2+
  \tfrac{35}{6}\beta_1^2\beta_0 \right)
  \ln^3\left(\frac{\mu_0^2}{\mu^2}\right) \right.
\nonumber \\
  &  \hspace{18em} \left.\null+\tfrac{7}{2} \left(\beta_3\beta_0
  +\beta_2\beta_1 \strut\right) \ln^2\left(\frac{\mu_0^2}{\mu^2}\right)
  +\beta_4 \ln \left(\frac{\mu_0^2}{\mu^2}\right)  \right]
  a_{\mu_0}^6 +\mathcal{O}(a_{\mu_0}^7).
\end{align}
\end{widetext}
where $a_\mu\equiv \alpha_s(\mu)/(4\pi)$; i.e.\ the normalized coupling determined at the scale $\mu$, can be related to its measured value at the initial scale $\mu_0$. The main differences between the two solutions are that the first is the direct solution of Eq.~\eqref{betaf} parameterized with the $\Lambda$ scale parameter, while the second is the same solution, but referred directly to the measured value of the coupling at a given scale $\alpha_s(\mu_0)$, expanded in $a_{\mu_0}\equiv \alpha_s(\mu_0)/(4\pi)$ and truncated at 5-loop accuracy, $\mathcal{O}(\alpha_s^7)$, as shown in Eq.~\eqref{5L-as}.

\section{Event-shape variables and the renormalization scale}
\label{sec:eventshapes}

A deeper knowledge of the geometrical structure of the process can be obtained considering event-shape variables, such as the thrust and $C$-parameter distributions. These observables are more exclusive with respect to the final state, as compared to decay rates, and are also particularly suited for the determination of the strong coupling $\alpha_s(M_Z)$~\cite{Kluth:2006bw}.

The thrust ($T$) and the $C$-parameter ($C$) are defined as:
\begin{align}
  T &= \max\limits_{\vec{n}}
  \left(\frac{\sum_{i}|\vec{p}_i\cdot\vec{n}|}{\sum_{i}|\vec{p}_i|}\right),
\\[1ex]
  C &= \frac{3}{2}
  \frac{\sum_{i,j}|\vec{p_i}||\vec{p_j}|\sin^2\theta_{ij}}
       {\left(\sum_i|\vec{p_i}|\strut\right)^2},
\end{align}
where $\vec{p}_i$ indicates the three-momentum of particle~$i$ and the sums run over all particles in the hadronic final state; $\vec{n}$ is a unit vector that is varied to maximize thrust $T$, determining the so-called thrust axis, i.e.\ $\vec{n}_T$. Alternatively, the variable $(1{-}T)$ is often used, which for 3-jet production at LO is restricted to the range $0<(1{-}T)<\frac13$. We have back-to-back and spherically symmetric events for $T=1$ and $T=\frac23$ respectively. For the $C$-parameter, $\theta_{ij}$ is the angle between $\vec{p_i}$ and~$\vec{p_j}$. For 3-jet production at LO the $C$-parameter is restricted by kinematics to the range $0\leq C \leq\frac34$.

Theoretical results for normalized infrared-safe observables,
such as the thrust or $C$-parameter distribution for the
annihilation process $e^+ e^-\rightarrow 3\,$jets at the $Z^0$ peak
to NNLO accuracy, are given in Refs.~\cite{DelDuca:2016ily, DelDuca:2016csb, Gehrmann-DeRidder:2014hxk, Gehrmann-DeRidder:2007nzq, GehrmannDeRidder:2007hr, Weinzierl:2008iv, Weinzierl:2009ms}.
When using the conventional scale-setting method, only one value of $\alpha_s$, at the kinematic scale $\sqrt{s}$ of the process, can be extracted with errors given by the uncertainty in choosing the renormalization scale, while using the PMC$_\infty$, which is based on intrinsic conformality (iCF), it is possible to determine the strong coupling at any scale, even up to the highest scale where the theory is expected to breakdown. As shown in Eq.~\eqref{5L-as}, the final renormalization scale $\mu$ is set by using the PMC$_\infty$ method while $\mu_0$ can be set to any value in the perturbative range; we can thus determine $\alpha_s(\mu_0)$ at any scale. Results for the PMC$_\infty$ for both thrust and $C$-parameter have been discussed in Refs.~\cite{DiGiustino:2023jiq, DiGiustino:2022ggl, DiGiustino:2020fbk, DiGiustino:2021nep}. We remark that the PMC$_\infty$ solves the renormalization scale ambiguity and the scales, related to each order of calculation, are fixed perturbatively by determining the $\beta_0$ term of each scale at each order of accuracy. These scales take into account the dynamics of the underlying quark and gluon subprocess and thus fix the magnitude of the strong coupling at each order and for the corresponding value of the selected variable $(C,1{-}T))$. In fact, these scales are not single valued but are themselves functions of the kinematic scale, $\sqrt{s}$, and of the event-shape variable via the coefficients of the number of flavors, $N_f$, related to the loop integrals and thus to their UV-divergent part. The PMC$_\infty$ scales are fixed order by order and the last scale is set to the kinematic scale of the process $\sqrt{s}\equiv M_Z$, given the missing knowledge of the higher-order $\beta_0$-term. We have defined this type of uncertainty as the second kind of residual scale dependence~\cite{Huang:2021hzr}. We remark that both the first and second kind of residual scale uncertainties are of the same order as the uncertainties in the first missing higher-order contribution and thus both have a mere perturbative nature linked to the perturbative series and its accuracy.
\\
\section{Fitting Method}
\label{sec:method}

  The aim of this analysis is to display and compare the results obtained in determining the strong coupling at the $Z^0$-mass scale, $\alpha_s(M_Z)$, by using the PMC$_\infty$ and the conventional scale-setting procedure, with a focus on the theoretical uncertainties given by the renormalization scale ambiguity. We thus adopt the method based on
the Maximum Likelihood and, in particular, we determine the value of the coupling parameter at the minimum of the $\chi^2$ variable. The $\chi^2$ fit is the result of minimizing
\begin{equation}
  \chi^2 \equiv \sum_{i}
  \left[
    \left( \langle y\rangle^{\mathrm{expt.}}_i 
          -\langle y\rangle^{\mathrm{th.}}_i \right)/\sigma_i 
  \right]^2,
  \label{eq:chi2def}
\end{equation}
where $i$ runs over all available bins, $\langle y\rangle^{\mathrm{expt.}}_i$ is  the experimental data  value, $\sigma_i$ the corresponding experimental uncertainty and $\langle y\rangle^{\mathrm{th.}}_i$ the theoretical prediction. The $\chi^2$ value is minimized with respect to $\alpha_s(M_Z)$ for the thrust and $C$-parameter. 

  The $\chi^2$ analysis  has been performed following the approach of Ref.~\cite{ALEPH:1996oqp}; thus, in the first approximation, considering the bins as being independent of each other and the experimental errors $\sigma_i$ determined by summing the statistical and systematic errors (due to detector effects and to hadronization) in quadrature. Effects due to the bin-to-bin correlation and to the intrinsic correlation among the event-shape results within one experiment, have been considered only in the final stage of the analysis, when all results are combined in the weighted average to obtain the highest precise determination of the strong coupling  $\alpha_s(M_Z)$.

In order to determine the experimental error on the fitted parameter, we use the standard definition of the error for this procedure, given by the upper and lower values of the parameter determined at $\chi^2_{min}+1$, while the theoretical errors have been determined by using the standard criteria; i.e.\ for the conventional scale setting by varying the renormalization scale in the range of two ($M_Z/2\leq\mu_r\leq2M_Z$) for the entire observable and for the PMC$_\infty$ by varying only the last scale at the NNLO order in the same range of two, since this is the only scale that is affected by the missing higher-order contributions. 

  This is, in fact, a merely conventional procedure to determine errors due to the renormalization scale ambiguity and it can give a first indication on the uncalculated higher orders.

The high-precision experimental data have been taken from LEP \cite{ALEPH:2003obs}; in particular, these data consist of $42$ experimentally measured values for thrust, $0<(1{-}T)<0.42$, divided into $42$ bins centered and with spacing $0.01$, while for the $C$-parameter we have $50$ experimental measurements in the range $0<C<1$ divided in $50$ bins centered and with spacing of $0.02$. In order to fit the theoretical distributions to the experimental data, for the case of the $C$-parameter we have to reduce the $100$-bin partition shown in Ref.~\cite{Weinzierl:2009ms} to $50$~bins. This reduction is made by averaging over two consecutive values of the distributions for the LO, NLO and NNLO coefficients.

For the theoretical predictions, we use the EVENT2 program~\cite{Catani:1996jh} to precisely calculate the perturbative coefficients at NLO and for the NNLO we use the results of Ref.~\cite{Weinzierl:2009ms}. We use the RunDec program~\cite{Chetyrkin:2000yt} to evaluate the running coupling in the $\overline{\rm MS}$ scheme and the perturbative formula Eq.~\eqref{5L-as} when indicated differently.

\subsection{Best-fitting region}
\label{sec:BFR}
We perform a further analysis by defining the \emph{best-fitting range} (BFR) the bin interval that probabilistically fits the data best according the $\chi^2$ distribution. 

  The purpose of this analysis is to improve or at least to encode the method used in Ref.~\cite{Dissertori:2009ik} of setting the fitting interval in the range where: ``\dots\  a  good perturbative description is available and where hadronization and detector corrections are below \dots'', in order to eliminate any possible human prejudice or uncertainty due to the choice of the bin-interval. For the case of the $(1{-}T)$ and $C$-parameter event-shape variables, we know well that in the central region of 3-jet production the perturbative calculations give the best fit to data. Besides, introduction of other corrections, such as nonperturbative corrections and hadronization over the entire range of the event shape, does not improve the theoretical predictions; on the contrary, much larger errors are introduced. Thus, it is a common practice to reduce the fitting region to the perturbative region in order to improve precision and statistics. Although we agree with this procedure, we must admit that it is not entirely consistent with the Maximum-Likelihood procedure adopted to fit the parameters. We are basically adding one degree of freedom -- the interval range -- and we are eliminating it subsequently by introducing some criteria, which might not be entirely consistent with the minimum $\chi^2$ analysis. Here, we now introduce a new procedure to select the BFR interval.

In order to determine the BFR, we perform the $\alpha_s(Q)$ fit by minimizing $\chi^2$ for each possible bin interval in the full range of available values for the selected variable; we then calculate the cumulative distribution function for each interval:
\begin{equation}\label{eq:CDF}
    P_{\chi^2}(\chi^2_{\mathrm{min}};n_{\mathrm{dof}})  \equiv \int^{\chi^2_{\mathrm{min}}}_0 d\chi^2 f(\chi^2;n_{\mathrm{dof}}),
\end{equation}
where $f(\chi^2;n_{\mathrm{dof}})$ is the $\chi^2$ probability density distribution for $n_{\mathrm{dof}}$ degrees of freedom. We define the BFR as the interval of bins with the lowest $P_{\chi^2}(\chi^2_{\mathrm{min}};n_{\mathrm{dof}})$. The bin interval determined by the lowest $P_{\chi^2}(\chi^2_{\mathrm{min}};n_{\mathrm{dof}})$ corresponds to the interval with the value of $\chi^2_{\mathrm{min}}$ that, most likely, we can consider the lowest according to its $\chi^2$ distribution. In fact, this $\chi^2_{\mathrm{min}}$ is the value that has the highest probability to be exceeded.

We point out that the lowest $\chi^2$ or the lowest reduced $\chi^2/n_{\mathrm{dof}}$ would not work in determining the best-fitting range. In fact, by varying the range of bins, one is also implicitly varying $n_{\mathrm{dof}}$. Firstly, $\chi^2$ alone does not take into account the dependence on the degrees of freedom; secondly, the reduced $\chi^2/n_{\mathrm{dof}}$ is linear in $\chi^2$ and thus leads to different results with respect to the probability $P_{\chi^2}$ at small $\chi^2\sim 0$. We also point out that this procedure is consistent with the Maximum-Likelihood strategy. In fact, on the one hand, we determine the most likely lowest $\chi^2_{\mathrm{min}}$ according to the $\chi^2$ distribution, consistent with a minimum-$\chi^2$ strategy; on the other hand, we can demonstrate that the values of the distribution of the fitted parameter for each bin interval, in the absence of particular sources of bias (such as an incorrect model or particular background noise in the calculations), would be in agreement with a normal distribution with mean value, median and mode corresponding to the value of the parameter determined by the minimum~$P_{\chi^2}$.

\begin{figure}[htb]
  \centering
  \includegraphics[width=0.40\textwidth]{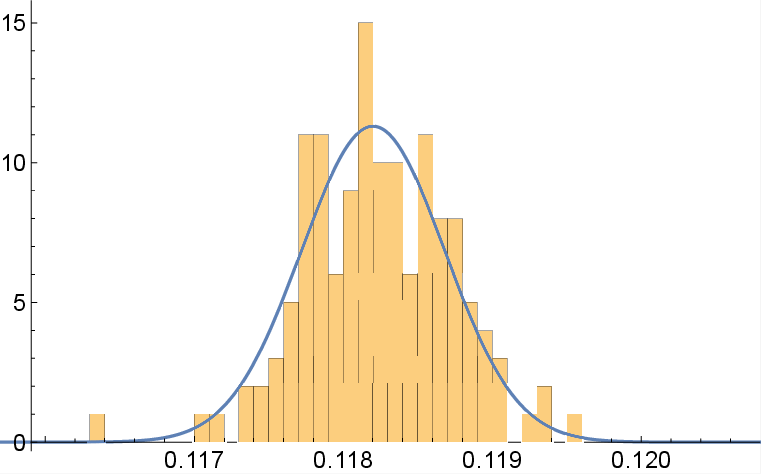}
  \caption{A histogram of the result sample for the fitted values of $\alpha_s(M_Z)$ for each possible interval in the range $0.38<C<0.72$ of the $C$-parameter, using PMC$_\infty$
  scale setting at $\sqrt{s}=M_Z$ center-of-mass energy.
  The normal distribution with mean $\overline{\alpha}_s(M_Z)=0.1182$ and variance $\sigma=0.0005$ is also shown. The curve is normalized to the area of the histogram.}
  \label{fig:normal-cpar}
\end{figure}

Results of the fits for $\alpha_s(M_Z)$, for each bin interval in the range $0.38<C<0.72$ (labelled as GFR) are displayed in Fig.~\ref{fig:normal-cpar} for the $C$-parameter distribution. In this range the mean, median and mode for this sample are $0.1182$ for the first two and $0.1181$ for the mode. To a good approximation, we can thus describe the results obtained with a normal distribution of mean $\overline{\alpha}_s(M_Z)=0.1182$ and variance $\sigma=0.0005$. We point out that while the minimum-$P_{\chi^2}$ strategy is applied to the entire range of values $0<C<1$, in order to isolate the normal behavior of the distribution, we must restrict the range to the perturbative region, where the theoretical predictions can be approximated by a linear model and the dominant source of error is statistics.

  Summing up, the second strategy may be applied once more information is known regarding the possible good-fitting range, i.e.\ in which range the errors are small and where the perturbative description is less affected by nonperturbative effects, as described by the first analysis presented in Ref.~\cite{Dissertori:2009ik}. We thus suggest that just setting the bin interval randomly or by guessing it without maintaining consistency with the Maximum-Likelihood strategy in the perturbative region introduces another error, which can be determined by the variance, $\sigma=0.0005$ shown above, of the sample consisting of all the possible intervals in the perturbative region.

\section{Results}
\label{sec:results}
Results for the strong coupling extracted from thrust are shown in Tables~\ref{tab:T_PMC_RUNDEC}, \ref{tab:T_PMC_PERT}, \ref{tab:T_CSS_RUNDEC} and from the $C$-parameter in Tables~\ref{tab:C_PMC_RUNDEC}, \ref{tab:C_PMC_PERT}, \ref{tab:C_CSS_RUNDEC}. We have labelled the different ranges (BFR), (TFR) and (GFR) to indicate respectively the Best Fitting Range, the Total Range of bins and the Range of bins used in Ref.~\cite{Dissertori:2009ik}.  In more detail, in the tables, we indicate with TFR the range of values $0.02<(1{-}T)<0.42$ for the thrust, excluding the first two bins, where the perturbative predictions need an improved description including resummation of the IR large logarithms and possibly also further nonperturbative or hadronization effects~\cite{Dasgupta:2003iq}. We underline that these corrections should be included also in the regions of the peak and in the multi-jet region above $(1{-}T)\simeq\frac13$. We point out that in the multi-jet region we have also the kinematic constraint at LO that cancels the LO contribution and we thus expect that the missing higher-order contributions would lead to an improvement~\cite{Abbate:2010xh} in this region. For the same reasons we restrict the $50$-bin sample of the $C$-parameter to $46$ neglecting the first two bins below $C<0.04$ and the two bins at the border $C\sim 0.75$, where there is an enhancement of logarithm behavior and theoretical predictions need to be improved with the inclusion of resummation and of nonperturbative effects~\cite{Hoang:2015hka}. We label this range of values with (TFR1*). We use the label (TFR2*) to indicate the sample of bins where three bins below $C<0.06$ have been excluded, leading to a sample of $45$ bins.

Results for the strong coupling at the $Z^0$ peak using the PMC$_\infty$, the 5-loop accuracy solution for the running coupling and the BFR method, are shown in Table~\ref{tab:T_PMC_RUNDEC} for thrust and Table~\ref{tab:C_PMC_RUNDEC} for the $C$-parameter; i.e.\ the values:
\begin{align}
  \alpha_s(M_Z) &= 
  0.1170^{+0.0007}_{-0.0007}(\mathrm{expt.})^{+0.0010}_{-0.0006}(\mathrm{th.}) 
  \nonumber \\
   & = 0.1170^{+0.0012}_{-0.0010},
\end{align}
with $\chi^2/n_{\mathrm{dof}}=0.80/11$ for thrust, and
\begin{align}
  \alpha_s(M_Z) &= 
  0.1181^{+0.0006}_{-0.0006}(\mathrm{expt.})^{+0.0011}_{-0.0007}(\mathrm{th.}) 
  \nonumber \\
  & = 0.1181^{+0.0013}_{-0.0009},
\end{align}
with $\chi^2/n_{\mathrm{dof}} =2.63/16$ for the $C$-parameter distribution, where the first error (expt.\@) is the experimental uncertainty and the second error (th.\@) is the theoretical uncertainty. As shown in the tables, the best fitting range (BFR) for the thrust and for the $C$-parameter agrees with the region that we retain are mainly perturbative and less affected by nonperturbative effects or by the IR logarithms, which here are not enhanced.
\begin{figure*}[htb]
  \includegraphics[height=0.35\textheight, width=0.49\textwidth]{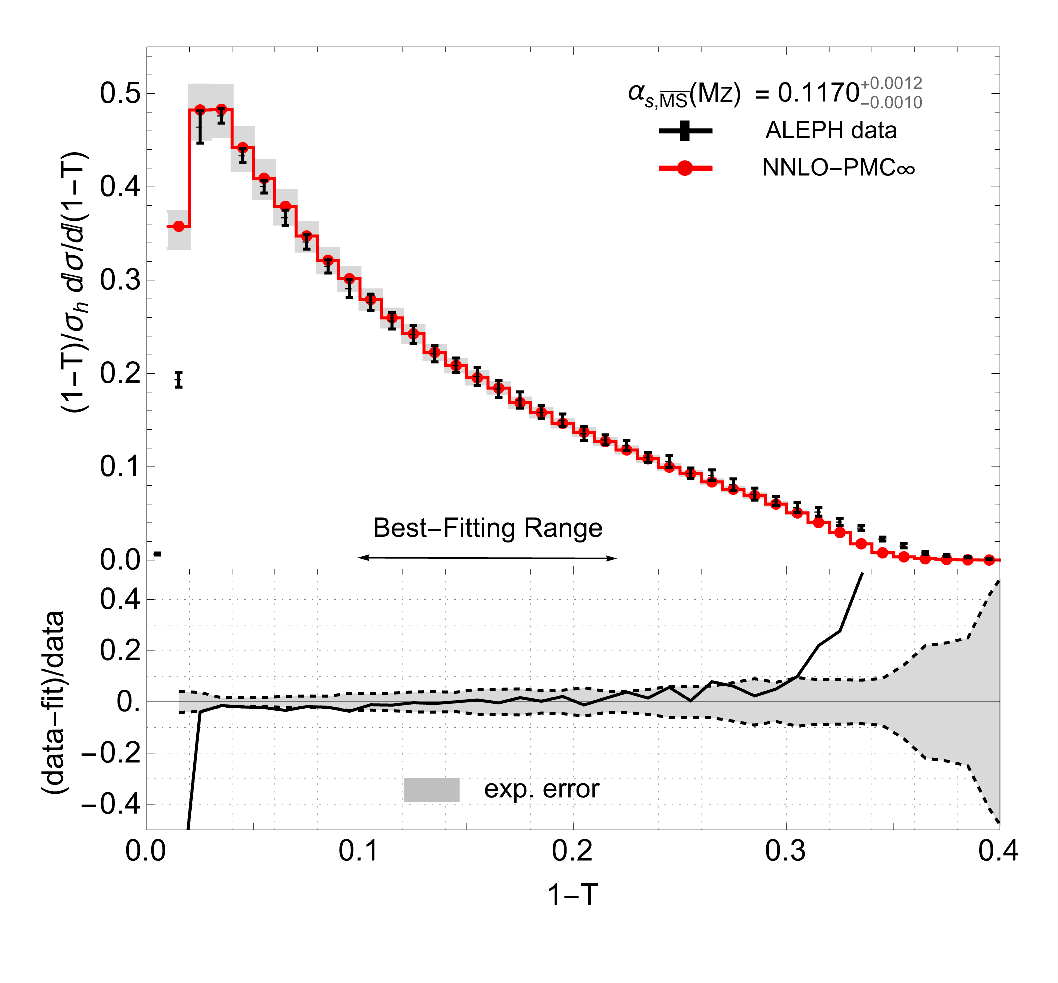},
  \includegraphics[height=0.35\textheight, width=0.49\textwidth]{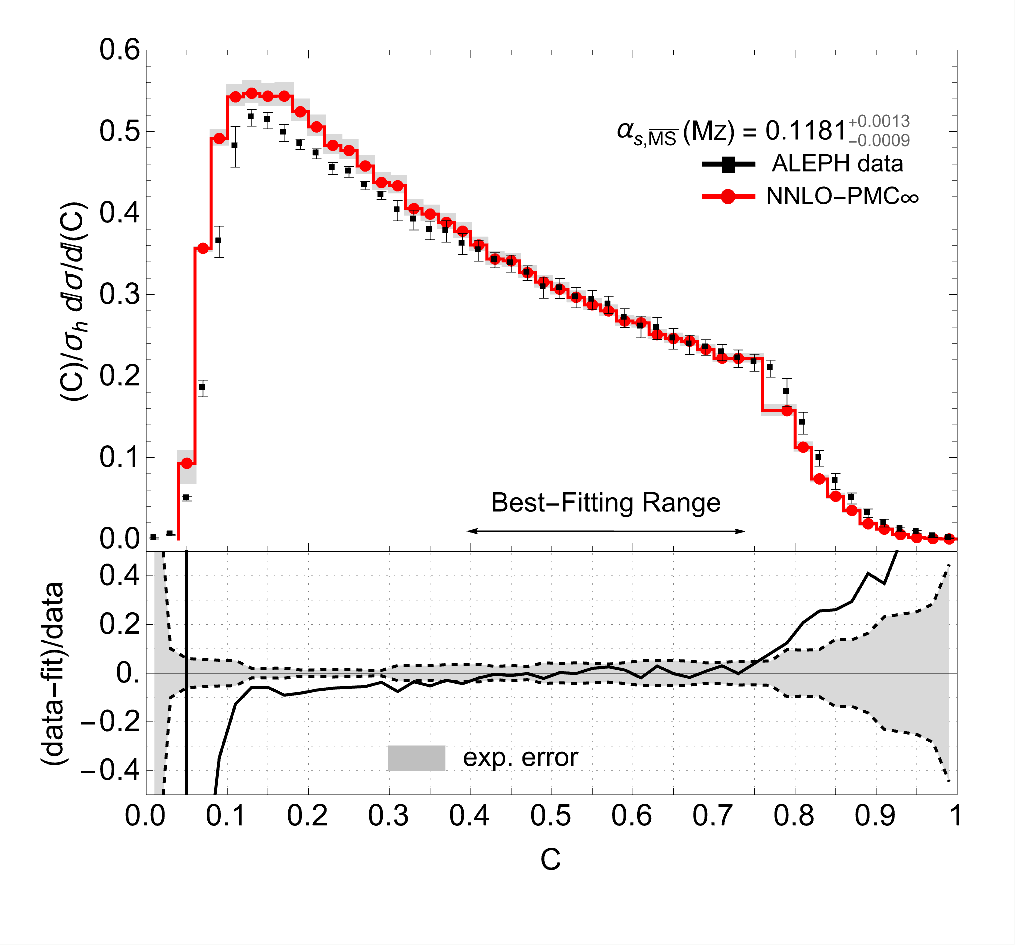}
  \caption{Comparison of the thrust and $C$-parameter distributions to  ALEPH data~\cite{ALEPH:2003obs}, using the PMC$_\infty$ scale setting and the values of $\alpha_s(M_Z)$ extracted in the best-fitting range. The shaded areas represent: upper) the theoretical errors; lower) the experimental errors.}
  \label{fig:fits}
\end{figure*}

Theoretical predictions for the thrust and $C$-parameter distributions using the extracted values of the strong coupling are shown in Fig.~\ref{fig:fits}, where the shaded areas represent in the upper-planes, the theoretical errors calculated using standard criteria of varying the unfixed scale in a range of two $(M_Z/2\leq \mu_r \leq 2 M_Z)$ and in the lower-planes the total relative experimental error, i.e. the $\sigma_i/(\langle y\rangle_i^{\rm exp})$  of Eq. \ref{eq:chi2def}.

  For a more detailed discussion regarding the PMC$_\infty$ thrust and $C$-parameter distributions, see e.g.\ Refs.~\cite{DiGiustino:2023jiq, DiGiustino:2020fbk, DiGiustino:2021nep}. We note that the BFR interval using the PMC$_\infty$ is in agreement with the interval chosen in Ref.~\cite{Dissertori:2009ik}, where detector effects and hadronization errors are below $25\%$ and the perturbative prediction are less affected by nonperturbative effects. In particular, in this region (i.e.\ the BFR), we find the best agreement between theory and experiment, as shown in the lower graph of Fig.~\ref{fig:fits}.

We underline that a different choice of the $\mu_0$ scale would lead to the same results for the strong coupling and would match perfectly the results with the same errors for the running coupling at the $Z^0$ peak, $\alpha_s(M_Z)$. We have performed the $\chi^2$ fit for the strong coupling at the Planck scale $\mu_0 \sim M_{\mathrm{Pl}} =1.22 \cdot 10^{19}$\,GeV, in the (BFR) using the PMC$_\infty$ and we have obtained the result: 

\begin{equation} 
  \alpha_s(M_{\rm Pl}) = 0.018837^{+0.000032}_{-0.000024}
\end{equation} 
with the same statistics $\chi^2/n_{\mathrm{dof}}$ and $P_{\chi^2}(\chi^2_{\mathrm{min}};n_{\mathrm{dof}})$, i.e.\ $0.8/11$ and $1.6\cdot 10^{-5}$ respectively. These results lead to the same predictions for $\alpha_s(M_Z)$ shown in Table~\ref{tab:T_PMC_RUNDEC}. 
We remark that in this approach, the physical scale is given by the final renormalization scale $\mu_{\rm PMC}$ set by using the PMC$_\infty$ method, while the initial scale, $\mu_0$, is purely conventional and is related to the final renormalization scale by renormalization group transformations.
Thus, the determination of the coupling at the Planck scale $ \alpha_s(M_{\rm Pl})$,  should not be considered as a direct measure of the coupling at the physical scale $M_{\rm Pl}$, but only as a test of the PMC$_\infty$ method.

Analogous results are obtained in the other bin ranges. We note that the results shown in Table~\ref{tab:T_PMC_RUNDEC}, for thrust using the PMC$_\infty$ and the 5-loop solution for the running coupling have very small differences with respect to the particular range and that the same prediction is obtained for the coupling using the 5-loop perturbative solution of Eq.~\eqref{5L-as}, as shown in Table~\ref{tab:T_PMC_PERT} in the BFR range. It is to be noted that in the case of thrust results for extracted values of the coupling using the PMC$_\infty$ in the BFR and in the total range of bins (TFR) are consistent within the errors. While for the $C$-parameter we find a significant reduction of the errors, of one order of magnitude in the range TFR1 using the PMC$_\infty$, independently from the particular solution used for the coupling, as shown in Table~\ref{tab:C_PMC_RUNDEC} and Table~\ref{tab:T_PMC_PERT}. This effect is possibly due to an effect of constraint of the particular bin at $C\sim 0.05$. In fact, in the range of bins TFR2 that excludes this particular bin, the effect disappear. Given the poor statistics, we would tend to exclude this prediction, but in a global fit that includes also nonperturbative effects over the entire range of values (TFR1) of the $C$-parameter and that would improve the goodness of the fit in this range, we would expect a higher precision in the determination of the strong coupling. We show results using the conventional scale setting for thrust and the $C$-parameter in Table~\ref{tab:T_CSS_RUNDEC} and Table~\ref{tab:C_CSS_RUNDEC} respectively. Comparing the results to those shown in Ref.~\cite{Dissertori:2009ik}, we note that, by using the conventional scale setting, we obtain the same errors for the strong coupling extracted from the NNLO theoretical predictions of thrust and the $C$-parameter. The extracted values of the strong coupling are analogous to those of Ref.~\cite{Dissertori:2009ik}; although in the case of the $C$-parameter we find a small deviation. This is possibly due to the small differences in the NNLO coefficient that have also been shown in Ref.~\cite{Weinzierl:2009ms}. Comparison between conventional scale setting and PMC$_\infty$ results clearly shows that PMC$_\infty$ leads to a significantly higher precision.

 A further analysis has been performed to investigate possible biassing effect due to some correlations in the BFR strategy. For this purpose we have kept held the BFR range of bins set using the ALEPH data and  then fitted the strong coupling at the $Z^0$ mass,  $\alpha_s(M_Z)$, using other independent data from the OPAL, DELPHI and L3 experiments~\cite{OPAL:2004wof, L3:2004cdh, DELPHI:2003yqh}. Results are shown in Table~\ref{tab:Tdatafits} and Table~\ref{tab:Cdatafits}.

  No particular bias effect due to the BFR approach is noted either for the central values or for the errors. The different central values obtained depend only on the experimental data used. Moreover, we find a reduced spread among the event-shape results according to the same experiment. The results appear to be in good agreement.

  In order to determine the strong coupling, we have first corrected the experimental errors  obtained from the minimum-$\chi^2$-fit for each event-shape measurement, by introducing the effects of the bin-to-bin correlation. The corrected errors are indicated as: Expt. Err.$^*$ in Tables: \ref{tab:Tdatafits} and \ref{tab:Cdatafits}. In particular, a robust estimation of the covariance matrix and of the bin-to-bin correlation parameter has been performed by using the same strategy as indicated in Ref.  \cite{Schmelling:1994pz} and in Ref. \cite{DELPHI:2000uri}, that suggests to set the correlation parameter $\rho_{\rm eff}$ in order to obtain a consistent $\chi^2$: i.e. $\chi^2(\rho_{\rm eff}) \equiv n_{\rm dof} $, when the $\chi^2_{\rm min}< n_{\rm dof}$. On the other hand, when $\chi^2_{\rm min}> n_{\rm dof}$, we rescale the errors until $\chi^2= n_{\rm dof}$ is satisfied. More in detail, we introduce the positive correlation for thrust and C-parameter results from ALEPH and for thrust results from the L3 experiment.\footnote{   We point out that this approach has some limitations, given by the assumption that the entire statistical gap is compensated with correlation.}

  Thus, combining and performing the weighted average of these independent values, using the asymmetric quadratic model for the errors shown in Ref.~\cite{Barlow:2003xcj} and including also the correlation between the event-shape variables thrust and $C$-parameter using a correlation parameter of $70\%$ (as shown in Refs.~\cite{Dissertori:2009ik, ALEPH:2003obs}), we obtain:  \begin{equation}
    \alpha_s(M_Z) = 0.1182^{+0.0007}_{-0.0007}. \label{eq:wa1}
  \end{equation}

\begin{figure}[htb]
  \centering
  \includegraphics[width=0.40\textwidth]{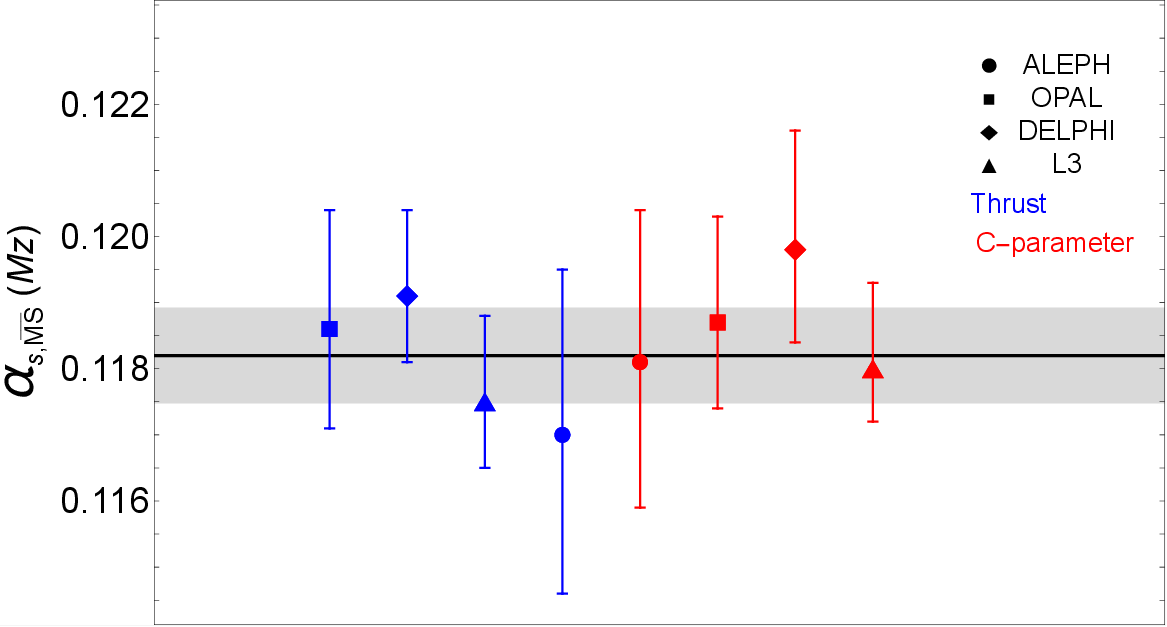} 
  \caption{Results for the strong coupling $\alpha_s(\mu)$ at the $Z^0$ peak for thrust and
  $C$-parameter in the BFR interval using ALEPH, OPAL, DELPHI and L3 \cite{ALEPH:2003obs,  OPAL:2004wof, DELPHI:2003yqh, L3:2004cdh}.
The shaded area shows the total errors for the average value $\alpha_s(M_Z)=
  0.1182^{+0.0007}_{-0.0007}$.} \label{fig:alphas}
\end{figure}

In Fig.~\ref{fig:alphas} we show the strong coupling $\alpha_s(M_Z)$ result determined by the weighted average and also the results for $\alpha_s(M_Z)$ from thrust and the $C$-parameter. Results for both thrust and the $C$-parameter are consistent with the world average~\cite{Navas:2024a1} and also with recent PMC applications presented in Refs.~\cite{Wang:2021tak,Wang:2019isi,Wang:2019ljl,Shen:2023qgz}.

   We remark that experimental errors have been obtained by first summing in quadrature statistical and  systematic errors ( related to detector and hadronization effects ) taken directly from the experimental data \cite{ALEPH:2003obs,  OPAL:2004wof, DELPHI:2003yqh, L3:2004cdh}) and subsequently correcting for bin-to-bin correlation or rescaling of the errors. We notice that in the highest-correlation case ( i.e. ALEPH),  we obtain $\rho_{\rm eff} \simeq 0.93$ for thrust and  $\simeq 0.83 $ for C-parameter), the corrected experimental errors obtained by introducing bin-to-bin correlations, (i.e. $\sim \pm 0,0023$ for thrust and  $\sim \pm 0,0020$ for the C-parameter)  agree with those obtained by summing the independent experimental errors related to statistics, detector and hadronization, shown in Refs.~\cite{Dissertori:2009ik, ALEPH:2003obs}, (i.e. $\sim \pm 0.0021$ ,$\sim \pm 0.0022$ for thrust and $\sim \pm 0.0018$ , $\sim \pm 0.0017$ for the C-parameter respectively). 
Note that these results come from the different bin-intervals chosen in Refs.~\cite{Dissertori:2009ik, ALEPH:2003obs}.
This indicates that in the worst case (that of very high correlation)
the BFR method can lead to errors that are no larger than the
errors obtained by a separate treatment of the statistical and
systematic errors, as shown in the previous analyses.
 An improved treatment of the errors (especially those coming from the hadronization models, e.g. including the full correlation matrices as obtained from the Monte Carlo simulations) would be necessary in order to test the BFR method and verify its results to high accuracy. 
  Thus, if we neglect the systematic errors for the time being, a direct comparison of the total error in Eq.\ref{eq:wa1}, with the total error obtained by summing in quadrature the statistical and theoretical errors obtained in Ref. \cite{Dissertori:2009ik}, (i.e. $\Delta \alpha_s (M_Z) \simeq \pm 0.0036$), shows that the PMC$_\infty$ method can lead to a clear improvement in precision with respect to the conventional scale setting adopted in Ref. \cite{Dissertori:2009ik}.

\subsection{Comment on the BFR method and results}

The BFR is designed to eliminate the errors arising from the bin-interval selection in the perturbative region, which have been neglected in the previous analyses. This method leads to a bin interval in which the theoretical perturbative predictions better
ﬁt the data. In this region non-perturbative effects and hadronization corrections are lower, in agreement with the range selected in Ref.~\cite{Dissertori:2009ik}. 
Moreover, the BFR strategy tends to minimize the discrepancy between theoretical predictions and data by determining the minimum $\chi^2_{\mathrm{min}}$,  and maximizing the $n_{\rm dof}$ the number of degrees of freedom, by the lowest $P_{\chi^2}(\chi^2_{\mathrm{min}};n_{\rm dof})$; no extra criteria are assumed to set the BFR interval. 
The lowest  $P_{\chi^2}(\chi^2_{\mathrm{min}};n_{\rm dof})$ leads to the interval, that with 
  the highest probability, we can consider the BFR leading to the best fit for the strong coupling in agreement with the Maximum-Likelihood strategy. 

This strategy is based on the assumption that all bins are independent and statistically valid and is recommended when the errors given by the assumption of a wrong theoretical model are larger than the statistical errors. When this condition is satisfied, there is, in principle, no risk of overfitting or of overestimating the errors or of biassing either the central values or errors. 
If the statistical fluctuations are comparable with the theoretical model discrepancies, there could be a risk of selection bias; this is particularly evident when the BFR-interval is, for instance, in the non-perturbative region and it could be subsequently discarded, but it would not be detectable when the BFR is in the perturbative region. 
However, we expect that the latter case would occur when a bin-interval with a low number of bins is selected (e.g. when $n_{\rm dof}<5$) leading to both poor statistics and a not-too-good $\chi^2_{\rm min}$ value. In general, given that the BFR tends to maximise the number of bins in the interval, we would expect that in the case of a good statistical sample, this would be quite rare to occur. 

\section{Conclusion}
\label{sec:conclusions}

We have shown in this article that by using the PMC$_\infty$ scale setting, one is able to improve significantly the precision of the fits of the strong coupling $\alpha_s(M_Z)$ with respect to the conventional method and the obtained results are consistent with the present values of the PDG~\cite{Navas:2024a1} and also with the previous applications of the PMC to the event-shape variables \cite{Wang:2021tak, Wang:2019isi,Wang:2019ljl}. Scale invariant predictions can be obtained using the PMC$_\infty$, since this method is based on the intrinsic conformality (iCF), leading to a complete independence of the results from the scale entering the extracted value of the strong coupling. This shows that the entire coupling function can be determined from a single experiment at a single scale.  By virtue of the iCF, this method allows the entire determination of the coupling at any scale and is thus essential for the high-precision unification of the QCD coupling over all domains~\cite{deTeramond:2024ikl}.

We have also proposed a new method to eliminate the uncertainties related to the choice of the interval in the perturbative region, which is consistent with the Maximum-Likelihood approach. The purpose of this method is to determine the best fitting region (BFR), i.e.\ the interval of bins in the perturbative region, that best fit the data among all possible intervals of bins. By applying the BFR method, the PMC$_\infty$ and using the strong coupling at 5-loop accuracy for thrust and the $C$-parameter, we have obtained results that are consistent with the perturbative predictions and the extracted values of the strong coupling at the $Z^0$ mass scale are consistent with the world average but with improved precision with respect to previous results.
For further investigations, we would recommend using the
BFR strategy as a reﬁning strategy, i.e. as a methodology to be
applied once there is sufficient knowledge of the theoretical
predictions and we also would recommend further investigation of
the method itself by improving either the treatment of systematic
errors and correlation effects or by performing a “look-elsewhere”
analysis in order to test the methodology and its results to the
highest degree.

\begin{acknowledgements}
  LDG thanks Matthias Steinhauser for helpful support on the RunDec code.
  XGW is supported in part by the Natural Science Foundation of China under Grant No.12175025 and No.12347101. 
SQW is supported in part by the Natural Science Foundation of China under Grant No. 12265011; by the Project of Guizhou Provincial Department under Grant No.YQK[2023]016 and No. ZK[2023]141. SJB is supported in part by the Department of Energy Contract No.~DE-AC02-76SF00515. SLAC-PUB-17782.
\end{acknowledgements} 

\begin{widetext}
\begin{table*}[htb]
\begin{tabular}{|c|c|c|c|c|c|c|c|}
  \hline
 \multirow{2}*{ Experiment}   &  \multirow{2}*{ $(1{-}T)$ -- BFR  }  &  
 \multirow{2}*{$\alpha_s(\mathbf{M_Z})$ }  & \multirow{2}*{Tot. Err.}  &   \multirow{2}*{Expt. Err.$^*$}  & \multirow{2}*{Th. Err.}   & \multirow{2}*{ $\chi^2_{\mathbf{min}}/n_{\mathrm{dof}}$}  & \multirow{2}*{ $P_{\chi^2_{\mathbf{min}}}^{n_{\mathrm{dof}}}$}  \\ & & & & & &  & \\ \hline
   ALEPH  & $(0.10 ; 0.22)$ &  $0.1170$   & $^{+0.0025}_{-0.0024}$  & $^{+0.0023}_{-0.0023}$  & $^{+0.0010}_{-0.0006}$  &  0.80/11 & $1.6\cdot 10^{-5}$   \\  
  \hline
   OPAL  &  $(0.10 ; 0.22)$ &  $0.1186$   & $^{+0.0018}_{-0.0015}$  & $^{+0.0014}_{-0.0014}$  & $^{+0.0012}_{-0.0007}$  &  5.0/2  & 0.92 \\  
  \hline
   DELPHI  & $(0.10 ; 0.22)$ &  $0.1191$   &  $^{+0.0013}_{-0.0010}$  & $^{+0.0007}_{-0.0007}$  & $^{+0.0011}_{-0.0007}$  & 21.9/5  & $0.9995$    \\  
  \hline
   L3  & $(0.10 ; 0.22)$   &  $0.1175$   &  $^{+0.0013}_{-0.0010}$   & $^{+0.0009}_{-0.0009}$   & $^{+0.0010}_{-0.0006}$  & 2.17/4  & 0.30  \\  
  \hline
\end{tabular}
\caption{Results for the value of the strong coupling $\alpha_s(M_Z)$ at 5-loop accuracy as implemented in the RunDec program, using the PMC$_\infty$ for thrust distribution at NNLO and using OPAL, DELPHI and L3 data \cite{OPAL:2004wof, L3:2004cdh, DELPHI:2003yqh} in the BFR fixed using ALEPH~\cite{ALEPH:2003obs}.}\label{tab:Tdatafits}
\end{table*}

\begin{table*}[htb]
\begin{tabular}{|c|c|c|c|c|c|c|c|}
  \hline
 \multirow{2}*{ Experiment}   &  \multirow{2}*{$C$ -- BFR}  &  
 \multirow{2}*{$\alpha_s(\mathbf{M_Z})$ }  & \multirow{2}*{Tot. Err.}  &   \multirow{2}*{Expt. Err.$^*$}  & \multirow{2}*{Th. Err.}   & \multirow{2}*{ $\chi^2_{\mathbf{min}}/n_{\mathrm{dof}}$}  & \multirow{2}*{ $P_{\chi^2_{\mathbf{min}}}^{n_{\mathrm{dof}}}$}  \\ & & & & & &  & \\ \hline
 ALEPH  &   $(0.40;0.74)$  &  $0.1181$   & $^{+0.0023}_{-0.0022}$ & $^{+0.0020}_{-0.0020}$ & $^{+0.0011}_{-0.0007}$ & 2.63/16 & $6.8\cdot 10^{-5}$ \\  
  \hline
   OPAL  &  $(0.40;0.74)$ &  $0.1187$   & $^{+0.0016}_{-0.0013}$ & $^{+0.0011}_{-0.0011}$  & $^{+0.0012}_{-0.0006}$  & 2.79/2 & $ 0.75$ \\  
  \hline
   DELPHI   & $(0.40;0.74)$ &  $0.1198$   & $^{+0.0018}_{-0.0014}$  & $^{+0.0010}_{-0.0010}$  & $^{+0.0015}_{-0.0009}$ & 94.6/8 & $\sim 1.0 $ \\  
  \hline
   L3  &  $(0.40;0.74)$ &  $0.1180$   & $^{+0.0013}_{-0.0008}$ & $^{+0.0006}_{-0.0006}$ & $^{+0.0011}_{-0.0006}$& 6.86/6 & 0.67 \\  
  \hline
\end{tabular}
\caption{Results for the value of the strong coupling $\alpha_s(M_Z)$ at 5-loop accuracy as implemented in the RunDec program, using the PMC$_\infty$ for  $C$-parameter distribution at NNLO and using OPAL, DELPHI and L3 data \cite{OPAL:2004wof, L3:2004cdh, DELPHI:2003yqh} in the BFR fixed using ALEPH~\cite{ALEPH:2003obs}.}\label{tab:Cdatafits}
\end{table*}

\begin{table*}[htb]
\begin{tabular}{|@{\quad}l|c|c|c|c|c|c|}
  \hline 
  \multicolumn{1}{|wc{45mm}|}{Range}  
  & $\alpha_s(\mathbf{M_Z})$ & Tot.~error & Th.~error &  Expt.~error & $\chi^2_{\mathbf{min}}/n_{\mathrm{dof}}$ & $P_{\chi^2} (\chi^2_{\mathbf{min}},n_{\mathrm{dof}})$ 
  \\\hline
  $0.10 < (1{-}T) < 0.22$ (BFR) & $0.1170$ & $^{+0.0012}_{-0.0010}$ & $^{+0.0010}_{-0.0006}$ & $^{+0.0007}_{-0.0007}$ & 0.80/11 & $1.6\cdot 10^{-5}$ 
  \\\hline
  $0.02 < (1{-}T) < 0.42$ (TFR) & $0.1167$ & $^{+0.0016}_{-0.0011}$ & $^{+0.0016}_{-0.0010}$ & $^{+0.0003}_{-0.0004}$ & $195.1/39$ & $\sim 1.0$        
  \\\hline
  $0.09 < (1{-}T) < 0.25$ (GFR) & $0.1170$ & $^{+0.0012}_{-0.0009}$ & $^{+0.0010}_{-0.0006}$ & $^{+0.0007}_{-0.0007}$ & $3.9/15$ &  $1.9\cdot 10^{-3}$
  \\\hline
\end{tabular}
\caption{Results according to different interval ranges for the value of the strong coupling $\alpha_s(M_Z)$ at 5-loop accuracy, as implemented in the RunDec program, using the PMC$_\infty$ for the Thrust, $(1{-}T)$-distribution, at NNLO.}\label{tab:T_PMC_RUNDEC}
\end{table*}

\begin{table*}[htb]
\begin{tabular}{|@{\quad}l|c|c|c|c|c|c|}
  \hline
  \multicolumn{1}{|wc{45mm}|}{Range}  
  & $\alpha_s(\mathbf{M_Z})$ & Tot.~error & Th.~error &  Expt.~error & $\chi^2_{\mathbf{min}}/n_{\mathrm{dof}}$ & $P_{\chi^2} (\chi^2_{\mathbf{min}},n_{\mathrm{dof}})$ 
  \\\hline
  $0.04 < (1{-}T) < 0.22$ (BFR) &  $0.1170$   & $^{+0.0018}_{-0.0011}$ & $^{+0.0017}_{-0.0010}$ & $^{+0.0005}_{-0.0004}$ & 2.0/17 & $2.9\cdot 10^{-6}$ 
  \\\hline
  $0.02 < (1{-}T)< 0.42$ (TFR) &   $0.1177$ & $^{+0.0017}_{-0.0012}$ & $^{+0.0017}_{-0.0011}$ & $^{+0.0003}_{-0.0003}$ &  $186.5/39$ & $\sim 1.0$      \\\hline
  $0.09 < (1{-}T) < 0.25$ (GFR) &   $0.1174$ & $^{+0.0013}_{-0.0009}$ & $^{+0.0011}_{-0.0006}$ & $^{+0.0007}_{-0.0006}$ & $3.0/15$ &  $4.2\cdot 10^{-4}$
  \\\hline
\end{tabular}
\caption{Results according to different intervals for the value of the strong coupling $\alpha_s(M_Z)$ at 5-loop accuracy, using the perturbative formula Eq.~\eqref{5L-as} and the PMC$_\infty$ for the Thrust, $(1{-}T)$-distribution, at NNLO.}\label{tab:T_PMC_PERT}
\end{table*}

\begin{table*}[htb]
\begin{tabular}{|@{\quad}l|c|c|c|c|c|c|}
  \hline
  \multicolumn{1}{|wc{45mm}|}{Range}  
  & $\alpha_s(\mathbf{M_Z})$ & Tot.~error & Th.~error &  Expt.~error & $\chi^2_{\mathbf{min}}/n_{\mathrm{dof}}$ & $P_{\chi^2} (\chi^2_{\mathbf{min}},n_{\mathrm{dof}})$ 
  \\\hline
  $0.09 < (1{-}T) < 0.26$ (BFR) &  $0.1285$   & $^{+0.0059}_{-0.0042}$ & $^{+0.0058}_{-0.0041}$ & $^{+0.0008}_{-0.0009}$ & 2.2/16 & $1.8\cdot 10^{-5}$ 
  \\\hline
  $0.02 < (1{-}T)< 0.42$ (TFR) &   $0.1327$ & $^{+0.0059}_{-0.0039}$ & $^{+0.0059}_{-0.0039}$ & $^{+0.0006}_{-0.0005}$ & $240.8/39$ & $\sim 1.0$ 
  \\\hline
  $0.09 < (1{-}T) < 0.25$ (GFR) &   $0.1285$ & $^{+0.0060}_{-0.0041}$ & $^{+0.0059}_{-0.0040}$ & $^{+0.0009}_{-0.0008}$ & $2.0/15$ &  $2.9\cdot 10^{-5}$
  \\\hline
\end{tabular}
\caption{Results according to different interval ranges for the value $\alpha_s(M_Z)$ of the strong
coupling at 5-loop accuracy as implemented in the RunDec
program, using the conventional scale setting for the Thrust, $(1{-}T)$-distribution, at NNLO.}\label{tab:T_CSS_RUNDEC}
\end{table*}

\begin{table*}[htb]
\begin{tabular}{|@{\qquad}l|c|c|c|c|c|c|}
  \hline
  \multicolumn{1}{|wc{45mm}|}{Range}  
  & $\alpha_s(\mathbf{M_Z})$ & Tot.~error & Th.~error &  Expt.~error & $\chi^2_{\mathbf{min}}/n_{\mathrm{dof}}$ & $P_{\chi^2} (\chi^2_{\mathbf{min}},n_{\mathrm{dof}})$ 
  \\\hline
  $0.40 <C< 0.74$ (BFR) &  $0.1181$   & $^{+0.0012}_{-0.0009}$ & $^{+0.0011}_{-0.0007}$ & $^{+0.0006}_{-0.0006}$ & 2.63/16 & $6.8\cdot 10^{-5}$ 
  \\\hline
  $0.04 <C< 1.0$ (TFR1*) &   $0.11858$ & $^{+0.00014}_{-0.00017}$ & $^{+0.00002}_{-0.00010}$ & $^{+0.00014}_{-0.00014}$ & $771.0/45$ & $\sim 1.0$
  \\\hline
  $0.06 <C< 1.0$ (TFR2*) &   $0.1156$ & $^{+0.0019}_{-0.0011}$ & $^{+0.0019}_{-0.0011}$ & $^{+0.0003}_{-0.0002}$ & $430.6/44$ & $\sim 1.0$        
  \\\hline
  $0.36 <C< 0.74$ (GFR) &   $0.1178$ & $^{+0.0013}_{-0.0009}$ & $^{+0.0012}_{-0.0007}$ & $^{+0.0006}_{-0.0006}$ & $4.5/18$ &  $5.5\cdot 10^{-4}$
  \\\hline
\end{tabular}
\caption{Results according to different interval ranges for the value of the strong coupling $\alpha_s(M_Z)$ at 5-loop accuracy as implemented in the RunDec program, using the PMC$_\infty$ for the $C$-parameter ($C$-distribution) at NNLO.}\label{tab:C_PMC_RUNDEC}
\end{table*}

\begin{table*}[htb]
\begin{tabular}{|@{\qquad}l|c|c|c|c|c|c|}
  \hline
  \multicolumn{1}{|wc{45mm}|}{Range}  
  & $\alpha_s(\mathbf{M_Z})$ & Tot.~error & Th.~error &  Expt.~error & $\chi^2_{\mathbf{min}}/n_{\mathrm{dof}}$ & $P_{\chi^2} (\chi^2_{\mathbf{min}},n_{\mathrm{dof}})$ \\
  \hline
  $0.40 <C< 0.74$ (BFR) &  $0.1185$   & $^{+0.0013}_{-0.0009}$ & $^{+0.0012}_{-0.0007}$ & $^{+0.0006}_{-0.0006}$ & 2.3/16 & $2.5\cdot 10^{-5}$ 
  \\\hline
  $0.04 <C< 1.0$ (TFR1*) &   $0.11967$ & $^{+0.00015}_{-0.00020}$ & $^{+0.00003}_{-0.00013}$ & $^{+0.00015}_{-0.00015}$ & $730.0/45$ & $\sim 1.0$
  \\\hline
  $0.06 <C< 1.0$ (TFR2*) &   $0.1167$ & $^{+0.0020}_{-0.0013}$ & $^{+0.0020}_{-0.0013}$ & $^{+0.0002}_{-0.0002}$ & $424.8/44$ & $\sim 1.0$        
  \\\hline
  $0.36 <C< 0.74$  (GFR) &   $0.1183$ & $^{+0.0013}_{-0.0010}$ & $^{+0.0012}_{-0.0008}$ & $^{+0.0006}_{-0.0006}$ & $3.6/18$ &  $4.8\cdot 10^{-5}$ 
  \\\hline
\end{tabular}
\caption{Results according to different intervals for the value of the strong coupling $\alpha_s(M_Z)$ at 5-loop accuracy, using the perturbative formula Eq.~\eqref{5L-as} and the PMC$_\infty$ for the $C$-parameter ($C$-distribution) at NNLO.}\label{tab:C_PMC_PERT}
\end{table*}

\begin{table*}[htbp]
\begin{tabular}{|@{\qquad}l|c|c|c|c|c|c|}
  \hline
  \multicolumn{1}{|wc{45mm}|}{Range}  
  & $\alpha_s(\mathbf{M_Z})$ & Tot.~error & Th.~error &  Expt.~error & $\chi^2_{\mathbf{min}}/n_{\mathrm{dof}}$ & $P_{\chi^2} (\chi^2_{\mathbf{min}},n_{\mathrm{dof}})$ 
  \\\hline
  $0.30 <C< 0.48$ (BFR) &  $0.1309$   & $^{+0.0059}_{-0.0040}$ & $^{+0.0058}_{-0.0039}$ & $^{+0.0009}_{-0.0009}$ & 0.2/8 & $2.0\cdot 10^{-6}$ 
  \\\hline
  $0.04 <C< 1.0$ (TFR1*) &   $0.1204$ & $^{+0.0053}_{-0.0029}$ & $^{+0.0053}_{-0.0029}$ & $^{+0.0004}_{-0.0005}$ & $14150.4/45$ & $\sim 1.0$
  \\\hline
  $0.06 <C< 1.0$ (TFR2*) &   $0.1325$ & $^{+0.0039}_{-0.0031}$ & $^{+0.0039}_{-0.0031}$ & $^{+0.0004}_{-0.0004}$ & $932.3/44$ & $\sim 1.0$        
  \\\hline
  $0.36 <C< 0.74$  (GFR) &   $0.1297$ & $^{+0.0049}_{-0.0041}$ & $^{+0.0048}_{-0.0040}$ & $^{+0.0008}_{-0.0007}$ & $4.2/18$ & $3.2\cdot 10^{-4}$ \\\hline
\end{tabular}
\caption{Results according to different interval ranges for the value of the strong coupling $\alpha_s(M_Z)$ at 5-loop accuracy as implemented in the RunDec program, using the conventional scale setting for the $C$-parameter ($C$-distribution) at NNLO.}\label{tab:C_CSS_RUNDEC}
\end{table*}

\end{widetext}

\end{document}